# Frequency Dependence of Diagonal Resistance in Fractional Quantum Hall Effect via Periodic Modulation of Magnetic Field

**Shosuke Sasaki**

Shizuoka Institute of Science and Technology, 2200-2 Toyosawa, Fukuroi, 437-8555, Japan

E-mail: sasaki@ns.sist.ac.jp

Energy spectrum of fractional quantum Hall (FQH) states is composed of single electron energy (Landau energy) neglecting the Coulomb interactions between electrons, classical Coulomb energy and the quantum energy via quantum transitions. Herein, the sum of the Landau energy and the classical Coulomb energy depends upon the value of the filling factor continuously. However, the quantum transition energy discontinuously depends upon the value of the filling factor. This discontinuity yields energy gaps in many stable FQH states. The energy gaps for specific filling factors produce the precise confinement of Hall resistance.

A new experiment is considered as follows; the magnetic strength is fixed to the value to confine the Hall resistance at the filling factor of 2/3 as an example. Moreover the magnetic modulation with frequency $f$ is applied to the system. The frequency dependence of the diagonal resistance is measured. Then, it is shown in this paper that the diagonal resistance varies drastically at some frequency value $f_0$. We clarify the following relation between this value $f_0$ and the magnetic strength width d$B$ of Hall plateaus as $f_0 = e\, \mathrm{d}B / (4\, \mathrm{Pi}\, m)$, where $-e$ is the charge of electron, Pi =3.141592, and $m$ is the mass of electron.

## 1. Introduction

Precise experiments have been carried out in ultra-high-mobility samples [1, 2], and then many local minima of diagonal resistivity $\rho_{xx}$ were discovered at filling factors $\nu$=3/8, 3/10, 4/11, 4/13, 5/13, 5/17, 6/17, and so on. These local minima are not expected in traditional theories [3-7]. Accordingly, several theorists have proposed their extended models, for example, Wojs et al. [8], Smet [9], and Pashitskii [10]. Thus, many interesting problems related to the fractional quantum Hall effect remain. We examine a new experiment for FQHE in this paper as follows: An oscillating magnetic field is applied to a fractional quantum Hall state in addition to the static magnetic field. In this case, the diagonal resistance $R_{xx}$ depends upon the frequency of the oscillation. We discuss this frequency dependence in detail in the next sections.

## 2. Gap or gapless mechanism in energy spectrum of FQHS

Experimental values of Hall resistance $R_H(v)$ are precisely equal to $h/(e^2v)$ for the filling factors $v = 2/3, 1/3, 1/5, 2/5, 3/5$, and so on, where $h$ is Planck's constant and $e$ is the electron charge. The deviation $(R_H(v) - h/(e^2v))/R_H(v)$ is about $3 \times 10^{-5}$ for $v = 2/3$ and about $2.3 \times 10^{-4}$ for $v = 2/5$ (see reference [12]). This precise confinement of Hall resistance means the existence of an energy gap in the energy spectrum for these filling factors. The energy gap is discussed by many theorists; for examples, in references [5, 11, 13]. The energy per electron $\varepsilon(v)$ versus filling factor $v$ is the sum of Landau energy, classical Coulomb energy, and quantum transition energy [11]. The Landau energy is the eigenenergy of a single electron, neglecting Coulomb interaction; it depends upon magnetic field strength $B$ as $\hbar eB/(2m)$, where $\hbar$ is $h/(2\pi)$, and $m$ is the electron mass. The sum of the Landau energy and the classical Coulomb energy continuously depends upon the value of the filling factor $v$. On the other hand, the quantum transition energy discontinuously depends upon the value of $v$. The schematic figures of the energy spectra near $v = 2/3$ and $v = 3/4$ are shown in Fig. 1 (see reference[13]).

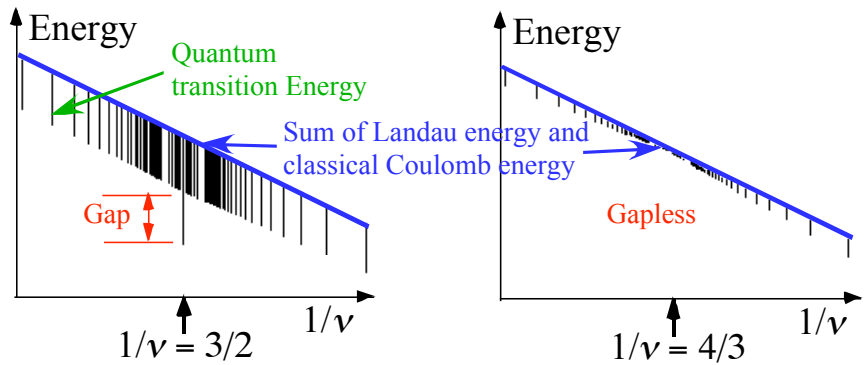

**Figure 1.** Energy spectra near $v = 2/3$ and $v = 3/4$

There is a gap at $v = 2/3$ and therefore a plateau appears in the curve of Hall resistance. On the other hand, FQH state with $v = 3/4$ has no energy gap, and therefore a plateau does not appear. This energy gap also yields vanishing of the diagonal resistance near $v = 2/3$. The diagonal resistance versus magnetic field strength has been measured in many experiments, one of which is shown in figure 2 described in the literature [1].

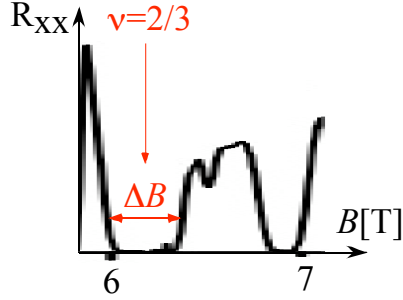

**Figure 2.** Experimental result of diagonal resistance $R_{xx}$ near $\nu = 2/3$ in reference [1]

As shown in figure 2, the diagonal resistance $R_{xx}$ is nearly equal to zero for $B_1 < B < B_2$, where the values of $B_1$ and $B_2$ are

$$B_1 \approx 6[T], \quad B_2 \approx 6.3[T] \text{ and } \Delta B = 0.3[T] \tag{1}$$

for the quantum Hall device used in reference [1]. Of course, these values depend on the Hall device. When the strength of magnetic field increases from $B_1$ to $B$, the energy $\varepsilon(\nu)$ increases as

$$\Delta\varepsilon = \hbar e(B - B_1)/(2m) \tag{2}$$

The FQH state with $\nu = 2/3$ is stable, when the value $\Delta\varepsilon$ is smaller than the energy gap $E_G(\nu)$. The FQH state becomes unstable when the magnetic strength $B$ becomes larger than $B_2$. According to the experimental data (see figure 2), the energy gap $E_G(\nu)$ is given as

$$E_G(\nu) = \hbar e(B_2 - B_1)/(2m) \quad . \tag{3}$$

Therefore, the numerical value is obtained as

$$E_G(2/3) \approx \hbar e(6.3 - 6)/(2m) \quad \text{for } \nu = 2/3, \tag{4}$$

for the Hall device used in reference [1].

### 3. Drastic increment of diagonal resistance

We consider a new experiment as:

(1) The magnetic field strength is fixed to maintain the FQH state with $\nu = 2/3$.

(2) A periodic magnetic modulation is added to the system. The frequency value is $f$.

(3) The diagonal resistance $R_{xx}$ is measured. Then the value of $R_{xx}$ depends upon the oscillation frequency $f$. The behaviour of $R_{xx}$ versus $f$ is predicted as shown in figure 3.

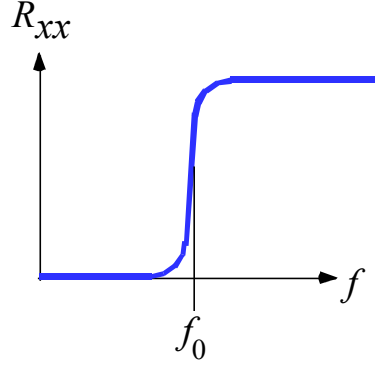

**Figure 3.** Diagonal resistance $R_{xx}$ versus $f$

The electrons at $\nu = 2/3$ cannot be excited for $fh < E_G$. Accordingly, the diagonal resistance $R_{xx}$ vanishes for $f < E_G/h$ at an ultra-low temperature. Therefore, the critical value $f_0$ is defined as

$$f_0 = E_G(\nu)/h \tag{5}$$

Accordingly, we obtain the frequency value by use of Eq. (3) as

$$f_0 = E_G(\nu)/(2\pi\hbar) = e(B_2 - B_1)/(4\pi\, m). \tag{6}$$

Electrons are excited when the frequency $f$ becomes larger than $f_0$. Then, $R_{xx}$ becomes very large. Consequently, the diagonal resistance value changes drastically near frequency $f_0$, as shown in figure 3. This critical frequency $f_0$ can be evaluated for the device used in reference [1]:

$$f_0 \approx 1.602\times 10^{-19}(0.3)/(4\pi\ 9.109\times 10^{-31}) \approx 4.2\times 10^9 \text{ for } \nu = 2/3 \tag{7}$$

That is to say, $f_0$ is about 4.2 GHz at $\nu = 2/3$. For the other filling factors, this critical frequency becomes a different value, e.g.

$$f_0 \approx 1.7\times 10^9 \text{ for } \nu = 3/5. \tag{8}$$

That is to say, $f_0$ is about 1.7 GHz at $\nu = 3/5$. Although the value of $f_0$ depends upon the configuration of a quantum Hall device, the relation between $f_0$ and $(B_2 - B_1)$ holds as in Eq. (6).

## 4. Conclusion

We have discussed a new phenomenon where the diagonal resistance changes drastically via magnetic modulation. The critical frequency value is related the width $(B_2 - B_1)$ as examined in this paper. The measurement of this critical frequency will give the precise value of the gap energy for each fractional filling factor with confinement of Hall resistance.

We can adopt a periodic modulation of electric current in a Hall device or an irradiation of electromagnetic wave if periodic modulation of magnetic field is difficult to apply in our new experiments.

The experiment proposed in this paper will clarify new properties of the fractional quantum Hall effect.